\documentclass[aps,pra,twocolumn,superscriptaddress]{revtex4-1}

\usepackage{booktabs}
\usepackage{color}
\usepackage[T1]{fontenc}
\usepackage{selinput}
\usepackage{changes}

\usepackage{graphicx}
\usepackage{dcolumn}
\usepackage{bm}
\usepackage{eqnarray}

\begin{document}


\title{Dispersive refraction of different light-to-heavy materials at MeV $\gamma$-ray energies}


\author{M. M. G\"unther}
\email[]{m.guenther@gsi.de}
\affiliation{Helmholtz-Institut Jena, Fr\"obelstieg 3, 07743 Jena, Germany}
\affiliation{GSI-Helmholtzzentrum f\"ur Schwerionenforschung GmbH, Planckstrasse 1, 64291 Darmstadt, Germany}
\author{A. V. Volotka}
\affiliation{Helmholtz-Institut Jena, Fr\"obelstieg 3, 07743 Jena, Germany}
\author{M. Jentschel}
\affiliation{Institut Laue-Langevin, 71 Rue des Martyrs, 38000 Grenoble, France}
\author{S. Fritzsche}
\affiliation{Helmholtz-Institut Jena, Fr\"obelstieg 3, 07743 Jena, Germany}
\affiliation{Theoretisch-Physikalisches Institut, Friedrich-Schiller-Universit\"at Jena, 07743 Jena, Germany}
\author{Th. St\"ohlker}
\affiliation{Helmholtz-Institut Jena, Fr\"obelstieg 3, 07743 Jena, Germany}
\affiliation{Institut f\"ur Optik und Quantenelektronik, Friedrich-Schiller-Universit\"at Jena, 07743 Jena, Germany}
\affiliation{GSI-Helmholtzzentrum f\"ur Schwerionenforschung GmbH, Planckstrasse 1, 64291 Darmstadt, Germany}
\author{P. G. Thirolf}
\affiliation{Ludwig-Maximilians-Universit\"at M\"unchen, Am Coulombwall 1, 85748 Garching, Germany}
\author{M. Zepf}
\affiliation{Helmholtz-Institut Jena, Fr\"obelstieg 3, 07743 Jena, Germany}
\affiliation{Institut f\"ur Optik und Quantenelektronik, Friedrich-Schiller-Universit\"at Jena, 07743 Jena, Germany}


\date{\today}

\begin{abstract}
The dispersive behavior of materials with atomic charge numbers varing from $Z = 4$ (beryllium, Be) to $Z = 82$ (lead, Pb) was investigated experimentally and theoretically at $\gamma$-ray energies up to 2 MeV. The experiment was performed at the double-crystal gamma spectrometer GAMS6 of the ILL in Grenoble. The experimental results were compared with theoretical calculations which account for all major elastic processes involved. Overall, we found a good agreement between theory and experiment. We find that for the development of refractive optics at $\gamma$-ray energies beyond those currently in use high-Z materials become increasingly attractive compared to the beryllium lens-stacks used at X-ray energies.
\end{abstract}

\pacs{}

\maketitle

\section{Introduction}
The investigation of basic optical properties of matter irradiated with visible light is of crucial importance for the development of optics in industrial as well as scientific applications. The discovery of X-ray radiation by Wilhelm-Konrad R\"ontgen \cite{Roentgen1898} and the subsequent development of X-ray radiation sources during the last century, opened a new field in optics and optical applications. The evolution towards modern brilliant X-ray sources such as synchrotrons made the investigation of the optical properties of materials up to about 100 keV possible. This allowed for the development of novel diffractive as well as refractive optics \cite{ice2011race,Snigirev,0022-3727-38-10A-042}. 

To date, the theoretical description the optical behaviour of materials has been validated experimentally in the X-ray energy regime from several hundreds of eV up to tens of keV. For $\gamma$-ray energies up to several MeV no experimental tests to verify the theory of the refractive index have been performed so far.

Knowledge on optical properties can be derived from the investigation of the forward elastic scattering processes. This allows for implementing different interaction processes via the complex forward scattering amplitude given by $A(E_{\gamma},0)= Re\, A(E_{\gamma},0)+i\, Im\, A(E_{\gamma},0)$. The complex forward scattering amplitude is related to the real as well as imaginary part of the complex refractive index, $n(E_{\gamma})=1+\delta (E_{\gamma})+i\beta (E_{\gamma})$, where the real part describes the phase shift between the incident wave and the scattered wave, and the imaginary part describes absorption. The real parts of the scattering amplitude and the refractive index are related in first order via the expresion
\begin{equation} \label{e0}
\delta(E_\gamma)=N_{\text{atom}}\frac{2\pi (\hbar c)^2}{E^2_{\gamma}}\,Re\, A(E_\gamma,0) \mbox{,}
\end{equation}
where $N_{\text{atom}}$ is the atom number density and $E_{\gamma}$ the photon energy. By the optical theorem,
\begin{equation} \label{e1a}
Im\, A(E_\gamma,0)=\frac{E_{\gamma}}{4\pi \hbar c}\sigma_{\text{tot}}(E_{\gamma}) \mbox{,}
\end{equation}
the imaginary part of the anomalous scattering factor is related to the total absorption cross section $\sigma_{\text{tot}}$.
The relation between the real and the imaginary part of the anomalous scattering factor is given by the Kramers-Kronig dispersion relation \cite{AlsNielsen2011,Toll},
\begin{equation} \label{e1}
Re\, A(E_\gamma,0)=\frac{2E_{\gamma}}{\pi}\mathcal{P}\int_{0}^{\infty}\frac{Im\, A(E',0)}{(E'^2 -E_{\gamma}^2)}dE' \mbox{.}
\end{equation}
Using this optical theorem, a direct relation between the phase shifting  and absorption processes can be obtained. This approach was used in a detailed investigation of several interaction processes of light with atoms by Toll in 1952 \cite{Toll}. Furthermore, by combining Eq.(\ref{e1}) with Eq.(\ref{e0}) and the optical theorem Eq.(\ref{e1a}), a direct link between the index of refraction and the scattering processes via the interaction cross section is reached. However, from a quantum physics point of view, it is mandatory that the scatterer is in a bound state, which can be excited by a dissipative process. Then, and only then, is this optical theorem applicable. 

Conversely, the measured optical properties, such as the refractive index, provide a sensitive observable for the investigation of the fundamental processes during the interaction of photons with matter and open the feasibility for tests of theoretical amplitudes at high photon energies. The Lorentz-Lorenz relation describes in general the scattering properties of an atom or compound material in relation with the macroscopic index of refraction:
\begin{equation} \label{e2}
n(E_\gamma) = 1 + 2\pi (\hbar c / E_\gamma)^2 \sum_k N_k A_k(E_\gamma,0)
 \mbox{,}
\end{equation}
where $N_k$ is the number of atoms of type $k$ per volume unit.
In the past, several experiments were performed for the investigation of the forward scattering amplitude within the X-ray regime \cite{PhysRevB.30.640,PhysRevB.42.1248,PhysRevA.35.3381}. The index of refraction was measured only up to 133 keV \cite{PhysRevA.91.033803}. Within the X-ray energy regime, when the incoming photon energies are much larger than the electron binding energy, the electrons can be treated as free electrons. This means that the energy dependent refractive index $n$ can be approximated by a coherent sum of forward amplitudes from each of the quasi-free electrons. The forward amplitude for light scattering on a free electron is given by the classical Thomson amplitude $-r_e$, where $r_e = e^2/mc^2$ is the classical electron radius. Within this approximation, which we refer later as \textit{classical approximation}, the real part of the complex scattering amplitude yields $Re A(E_\gamma,0) \approx -Z r_e$, where $Z$ is the atomic charge number. Accordingly, the real part of the refractive index reads in this approximation:
\begin{equation} \label{e3}
\delta(E_\gamma)=-Zr_e 2\pi \frac{(\hbar c)^2 N_A \rho}{E_{\gamma}^2 A} \mbox{,}
\end{equation}
for the case of pure materials. Here $\rho$ is the mass density, $N_A$ Avogadro's constant, and $A$ the molar mass.
From an application point of view, it is important to validate the classical approximation up to the $\gamma$-ray energy regime. To investigate the index of refraction at $\gamma$-ray energies it needs sophisticated experimental instrumentation, because the effect is expected to be very tiny. A first measurement of the $\gamma$-ray refractive index of silicon at $\gamma$-ray energies was performed by Habs et al. \cite{PhysRevLett.108.184802, PhysRevLett.118.169904}. Subsequent  experimental campaigns \cite{Guenther2017} showed that the conclusions drawn from the experimental results needed to be corrected at high energies due to a systematic error that had not been accounted for. A detailed discussion, including new results of the refractive index measurements of silicon, is presented in \cite{Guenther2017}.

In the present work we present the first detailed measurements of $\gamma$-ray refractive indices across a wide range of energies and atomic numbers  ranging from $Z=4$ to $Z=82$ at $\gamma$-ray energies up to 2 MeV. The aim was to investigate the dispersive behaviour as well as the $Z$-dependence of several materials with respect to the classical non-relativistic approximation. Furthermore, from an applied point of view, we were interested in the refractive properties for the development of refractive $\gamma$-ray optics. The measurements comprised for  the investigation of optical properties of a fluid as well as  of a compound material at such high photon energies.
In the following, we describe the experimental setup for measurements at the new high resolution gamma spectrometer GAMS6 at the Institut Max von Laue - Paul Langevin (ILL) in Grenoble (France). In the following chapters the results and the theory will be presented and discussed.

\section{Experimental setup}
The concept of the refractive index measurement is based on prism optics known since the dawn of modern optics: First, a collimated and monochromatic beam is prepared. Secondly, during the propagation through the prism surfaces, the beam is deviated due to refraction. Finally, the angular deviation with respect to the incident beam direction is measured via an analyzer device. Our experiments were performed at the high resolution double flat crystal gamma spectrometer GAMS6. The first crystal is used to collimate the gamma beam before propagation through the prisms. The second crystal is used to measure the angular deviation by rocking around Bragg's angle and scanning the intensity of the $\gamma$-ray line of interest (counted with a HPGe detector) as a function of the rocking angle. Both crystals are used in Laue diffraction mode and the result is a rocking curve according to the dynamical diffraction theory. The angular deviation due to refraction can be determined from the relative peak shift of the rocking curves with and without the prisms placed in the beam. The rotation angles of the crystals are measured using an optical angle interferometer (more details see \cite{KesslerJr2001187,PhysRevC.73.044303,Guenther2017}), operated in the visible range of the electro-magnetic spectrum via a frequency stabilized HeNe-Laser. The interferometer is of Mach-Zehnder type, providing a subnanoradian angular resolution for the measurement of both crystals. To minimize the impact of the refractive index of air, the spectrometer as well as the optical interferometers are operated in vacuum. The $\gamma$ beam is provided from an in-pile target placed close to the reactor core of the neutron high-flux reactor of the ILL. Via thermal neutron capture nuclear reactions in the target, prompt $\gamma$-ray emission is induced. As target we used $^{155,157}$Gd and $^{35}$Cl via the compound materials Gd$_2$O$_3$ and BaCl$_2$ as powder samples. The sample material, filled in graphite containers, was irradiated with a neutron flux of about 5$\times$10$^{14}$ s$^{-1}$ cm$^{-2}$, yielding a prompt $\gamma$-ray emission rate of about 10$^{17}$ s$^{-1}$. The sources were three containers of 2$\times$20 mm$^2$ placed at a distance of 22 m from GAMS6. The beam divergence is 10$^{-4}$ rad in the horizontal and 10$^{-3}$ rad in the vertical plane. The horizontal plane is the diffraction plane of the crystals. Via three different collimation systems the beam is collimated for the experiment. More details are described in \cite{Guenther2017}.

\begin{figure}
\includegraphics[width=0.47\textwidth]{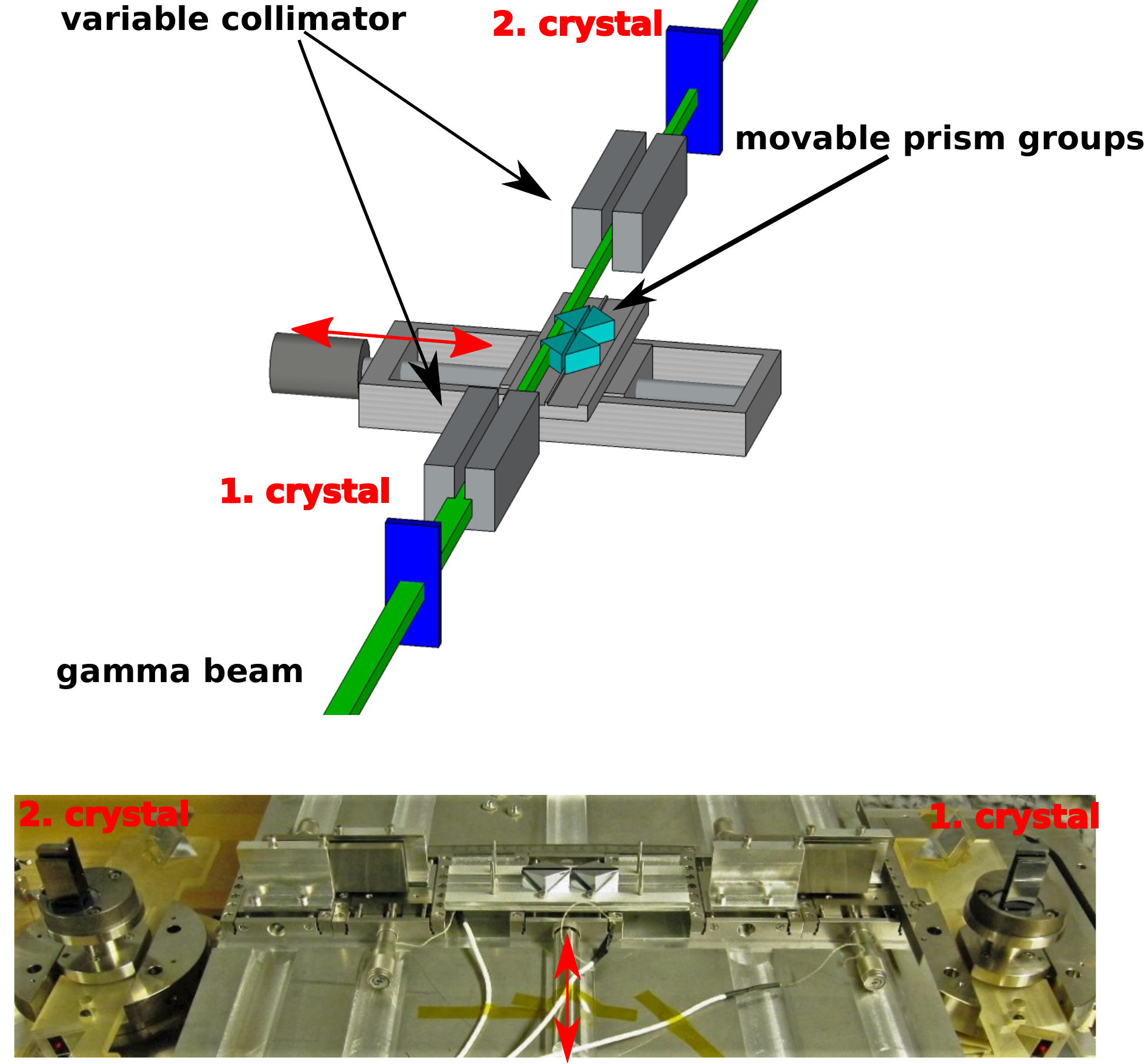}
\caption{\label{fig0}(color online) \textbf{Upper panel:} Schematic setup at GAMS6. The first crystal defines and collimates the $\gamma$-beam. A motorized translation stage moves the prism groups through the beam, and the second crystal allows to analyze the deviation due to refraction on the prism surfaces. \textbf{Lower panel:} Photograph of the setup. The $\gamma$ beam comes from the right. The crystals are mounted on the axis of a heterodyne Michelson interferometer, allowing for a high-resolution angular measurement.}
\end{figure}

A scheme of the setup for the refractive index measurement at GAMS6 is shown in Fig.\ref{fig0}. 
We used groups of prisms with the same geometry and a prism angle of 120 degrees, oppositely aligned. This allows for doubling the refraction effect and leads to a higher accuracy in the determination of the peak shift between the rocking curves of both, the right and the left prism orientations. The $\gamma$ beam was horizontally collimated to 3 mm and irradiated the prism surface. We used different prism materials to get a wide choice of atomic charge numbers $Z$. In case of fluid Hg we used hollow prisms, consisting of fused silica, filled with Hg. This method is similar to a historic experiment within the visible range of light, performed by Ernst Abbe more than one century ago in Jena (Germany) \cite{Abbe1874}. The prisms are mounted on a precision machined prism mount, which is placed on a motorized translation stage, allowing for an exact alignment and translation between the different orientations. An inter-crystal collimation system is used for an exact alignment of the prisms in the beam. Both, the prism system as well as the collimation system, are completely mechanically decoupled from the interferometer table for the angular measurement. Therefore, we did not impact the angular measurements due to vibrations during the translation movements. Furthermore, temperature sensors with 2 mK resolution were installed on sensible positions such as the interferometer, the crystals and the prisms to monitor the temperature variations during the movement of the motors. We observed no relevant temperature variations, requiring further corrections.

\begin{figure}
\includegraphics[width=0.47\textwidth]{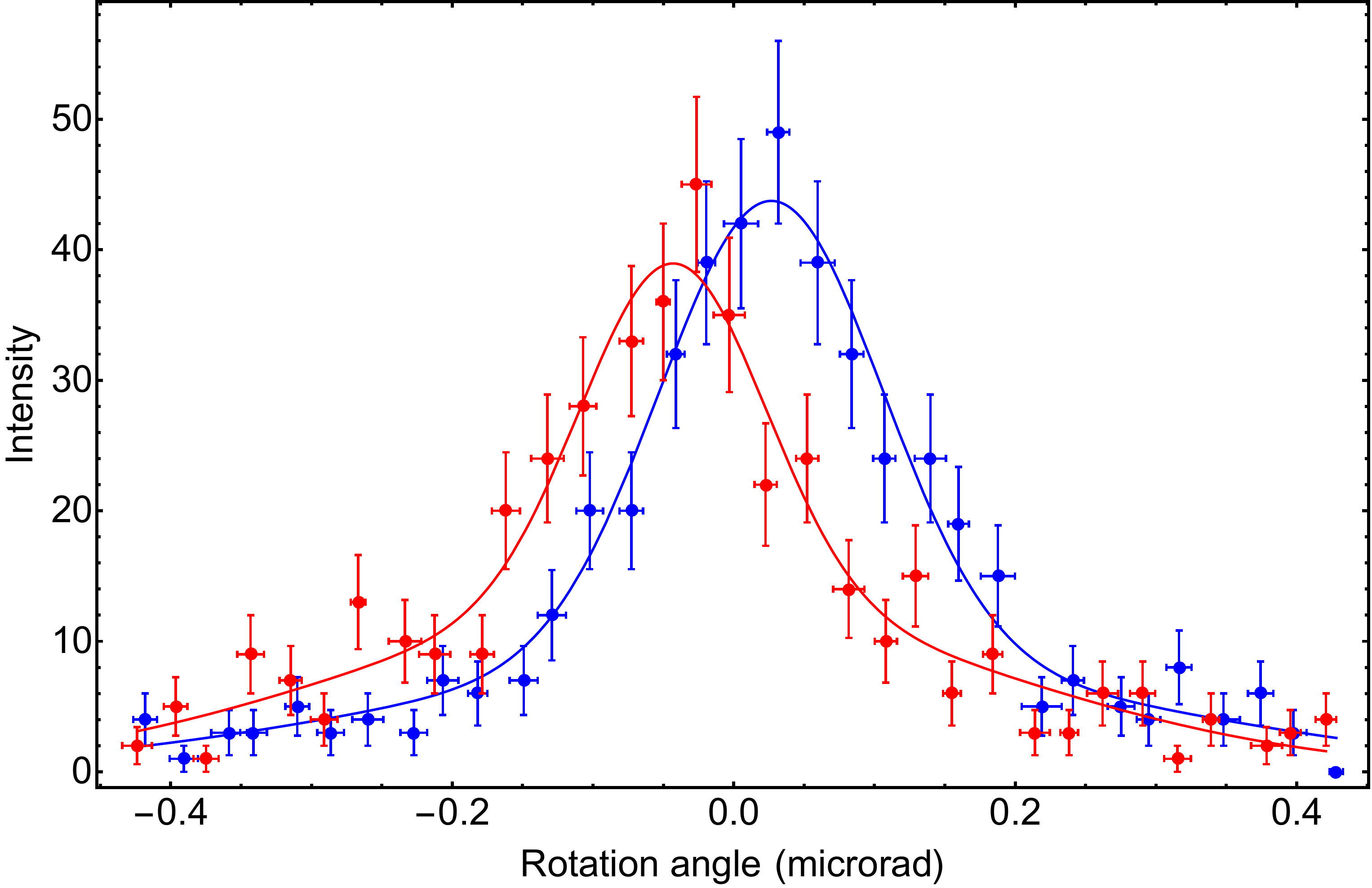}
\caption{\label{fig5}(color online) Rocking curves examplary from an up/down scan direction at 517 keV $\gamma$-ray energy in case of tantalum prisms. A double-Gaussian fit function (see solid lines; each with the same centroid) was applied to the measured data. For more details see text.}
\end{figure}
The measurement sequence of the refractive index measurement was the following: the oppositely oriented prism groups were moved into the collimated beam. For each position a rocking curve was scanned in up/down scan direction (see an example in Fig.\ref{fig5}). This yielded a set of a so-called 4-pack of scans allowing  any potential linear temporal drifts to be eliminated. For each material and each energy a series of 4-packs was recorded. Within each 4-pack the rocking curves were fitted by a double Gaussian fit function to obtain the peak position. This allowed within each 4-pack to extract one value of the deviation angle caused by refraction. Averaging all 4-pack values for a given energy and using the well known equation of prism optics, the corresponding value for the index of refraction is determined. Combining all values as function of the photon energy yields the dispersion curve for each material. A detailed principle description of the data analysis is given in \cite{Guenther2017}.

\section{Theory}
The optical relation Eq.(\ref{e0}) connects the real part of the refractive index $n(E_\gamma)$ with the real part of the complex forward scattering amplitude $A(E_\gamma,0)$, which can be expressed as the coherent superposition of amplitudes from various elastic scattering processes \cite{kane:1986:75}:
\begin{equation}
A(E_\gamma) = A_{\rm T}(E_\gamma) + A_{\rm R}(E_\gamma) + A_{\rm D}(E_\gamma) + A_{\rm NR}(E_\gamma)\,.
\end{equation}
Here, the angle argument in the amplitudes is dropped, since in the following we exclusively consider forward scattering. The amplitudes written above correspond to the nuclear Thomson (T), Rayleigh (R), Delbr\"uck (D), and nuclear resonance (NR) scattering channels. The nuclear Thomson scattering represents the scattering on a nucleus as a whole and its amplitude can be evaluated by the expression \cite{ericson:1973:604}
\begin{equation}
A_{\rm T}(E_\gamma) = - Z^2\frac{m}{M} r_e\,,
\end{equation}
where $M$ denotes the mass of a nucleus. By Rayleigh scattering we refer to the elastic scattering on bound electrons. In the nonrelativistic limit, when one can neglect the binding effects, the Rayleigh amplitude is given by
\begin{equation}
\label{e4}
A_{\rm R,\,class.}(E_\gamma) = - Z r_e\,,
\end{equation}
which is the classical Thomson scattering on $Z$ free electrons. To account for the relativistic corrections we use the following approach: For the photon energies under consideration, the Rayleigh scattering amplitude can be evaluated within an independent particle model \cite{volotka:2016:023418}, which means that the electron propagates in the mean field of the nucleus and all corresponding electrons. Within this approximation the Rayleigh amplitude can be calculated as a coherent sum of scattering amplitudes from each electron. The amplitudes for inner-shell electrons are evaluated within the second-order $S$-matrix approach, while for outer shells the modified form factor approximation is used. In particular, for the photon energy of 517 keV K and L shells and for 1164 keV and 1951 keV the K shell energy are calculated from the $S$-matrix approach. Such a prescription for the calculation of the Rayleigh scattering was proposed in Ref.~\cite{kissel:1980:1970}.
In Table~\ref{table0} we compare the relativistic correction to the Rayleigh scattering amplitude with results of previous studies. In the case of Zn ($Z = 30$), the scattering on the K shell only is treated by the $S$ matrix similar to Ref.~\cite{johnson:1976:692}, while for Pb ($Z = 82$) the K and L shells are included in the $S$-matrix approach just as in Refs.~\cite{johnson:1976:692,kissel:1980:1970}. The contributions of other electrons are accounted within the modified form factor approximation \cite{Franz}. For consistency, we add these contributions also to the values of Ref.~\cite{johnson:1976:692}. As one can conclude from the table, our results are in an excellent agreement with those of Refs.~\cite{johnson:1976:692, kissel:1980:1970}. Moreover, in the last two columns we present the results of the modified form factor approximation applied to all electrons as well as the predictions obtained by Toll's formula \cite{Toll}, which was derived using the dispersion relation to Sauter's approach \cite{Sauter1931}. Comparing the results of the modified form factor approximation and more accurate $S$-matrix formalism one can conclude that the first one provides quite good values and agreement rapidly improves for larger photon energies. In contrast, the values of the relativistic correction obtained by Toll's formula \cite{Toll} disagree strongly with all other predictions and, therefore, this suggests that there is a mistake in the formula, which to our knowledge was not yet discussed or  mentioned in the literature.
\begin{table}[]
  \centering
  \caption{\label{table0} Relativistic correction to the real part of the
  Rayleigh scattering amplitude $A_{\rm R}(E_\gamma)$, in units of $r_e$.
  The $S$-matrix calculations (this work; Johnson and Cheng
  \cite{johnson:1976:692}; Kissel, Pratt, and Roy \cite{kissel:1980:1970})
  are compared with the results of the modified form factor (MFF) and
  dispersion relation (DR) approximations.}
\begin{ruledtabular}  
\begin{tabular}{ccccccc}
$Z$ & $E_\gamma$, keV & $S$, TW
                             & $S$, \cite{johnson:1976:692}
                                    & $S$, \cite{kissel:1980:1970}
                                          & MFF  &   DR   \\ \hline
 30 &             279 & 0.08 & 0.08 &     & 0.09 &$-$0.03 \\
 30 &             662 & 0.09 & 0.09 &     & 0.09 &   0.01 \\
 82 &             412 & 0.88 & 0.89 & 0.9 & 0.99 &$-$1.44 \\
 82 &             889 & 0.98 & 0.98 & 1.0 & 0.99 &   3.71 \\ \hline
\end{tabular}
\end{ruledtabular}
\end{table}

The Delbr\"uck scattering amplitude in the first-order Born approximation was evaluated from the fourth-order vacuum polarization tensor in Ref.~\cite{papatzacos:1975:206} or employing the dispersion relation approach in Ref.~\cite{detollis:1977:499}. A simple fitting formula for the Delbr\"uck amplitude and valid for incident photon energies $E_\gamma < 4 mc^2$ was obtained in Ref.~\cite{solberg:1995:359},
\begin{eqnarray}\nonumber
A_{\rm D}(E_\gamma) = (\alpha Z)^2 \left [ 3.1735 \times 10^{-2} ( E_\gamma / mc^2 )^2 \right. \\ \nonumber
                                        + 3.1610 \times 10^{-4} ( E_\gamma / mc^2 )^4 \\
                                      \left.  + 1.4790 \times 10^{-5} ( E_\gamma / mc^2 )^6 \right ] r_e\,.
\end{eqnarray}
The Coulomb and screening corrections to the Delbr\"uck amplitude in the forward direction were evaluated in Refs.~\cite{overbo:1977:412,overbo:1979:299}. However, in the current study we neglect these contributions due to their smallness for the considered photon energies. Finally, photons can be elastically scattered via nuclear excitations. In the present experiment the energy linewidth of incoming photon beams was about 1 - 3 eV, which is several orders of magnitude smaller than the nuclear level spacing. Therefore, the nuclear excitation probability is negligible and we consider here only the scattering on the tails of the giant dipole resonances. The nuclear-resonance amplitude evaluated from the photonuclear absorption cross section is given by \cite{rullhusen:1981:1375}
\begin{equation}
A_{\rm NR,\,GDR}(E_\gamma) = \frac{E_\gamma^2}{4 \pi \hbar c r_e} \sum_j \sigma_j \Gamma_j
                             \frac{E_j^2-E_\gamma^2}{(E_j^2-E_\gamma^2)^2+E_\gamma^2 \Gamma_j^2}\,,
\end{equation}
where $j$ numerates the Lorentzian curves, which fit the photonuclear absorption data, $E_j$, $\Gamma_j$, and $\sigma_j$ are the position, width, and peak nuclear photo-absorption cross section of the $j$th resonance. The values of these parameters for the considered nuclei were taken from Ref.~\cite{atlas:1999}.

\section{Results and discussion}
\begin{figure*}
\includegraphics[width=1.\textwidth]{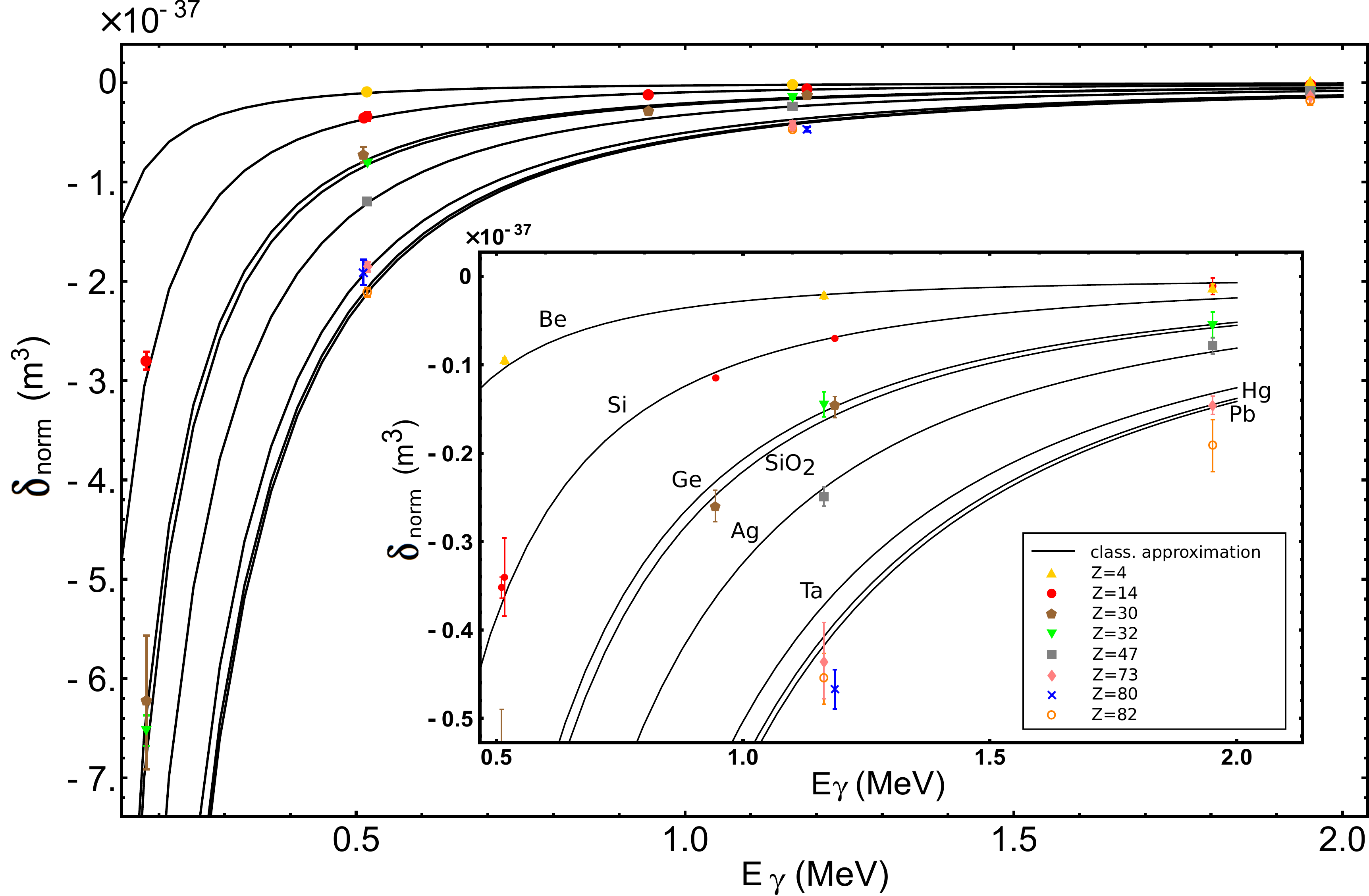}
\caption{\label{fig1}(color online) Dispersion curves of several materials, where the real part of the refractive index is normalized to the atom number density ($\delta_{\rm norm} = \delta / N_{\rm atom}$). The dots represent the measured $\delta_{\rm norm}$ for different energies $E_{\gamma}$ and materials. The inserted diagram shows in detail the deviation of $\delta_{\rm norm}$ from the classical approximation (black solid lines) for the high-$Z$ materials and at high photon energies from 500 keV up to about 2 MeV. For more details see text.}
\end{figure*}

\begin{figure*}
\includegraphics[width=0.8\textwidth]{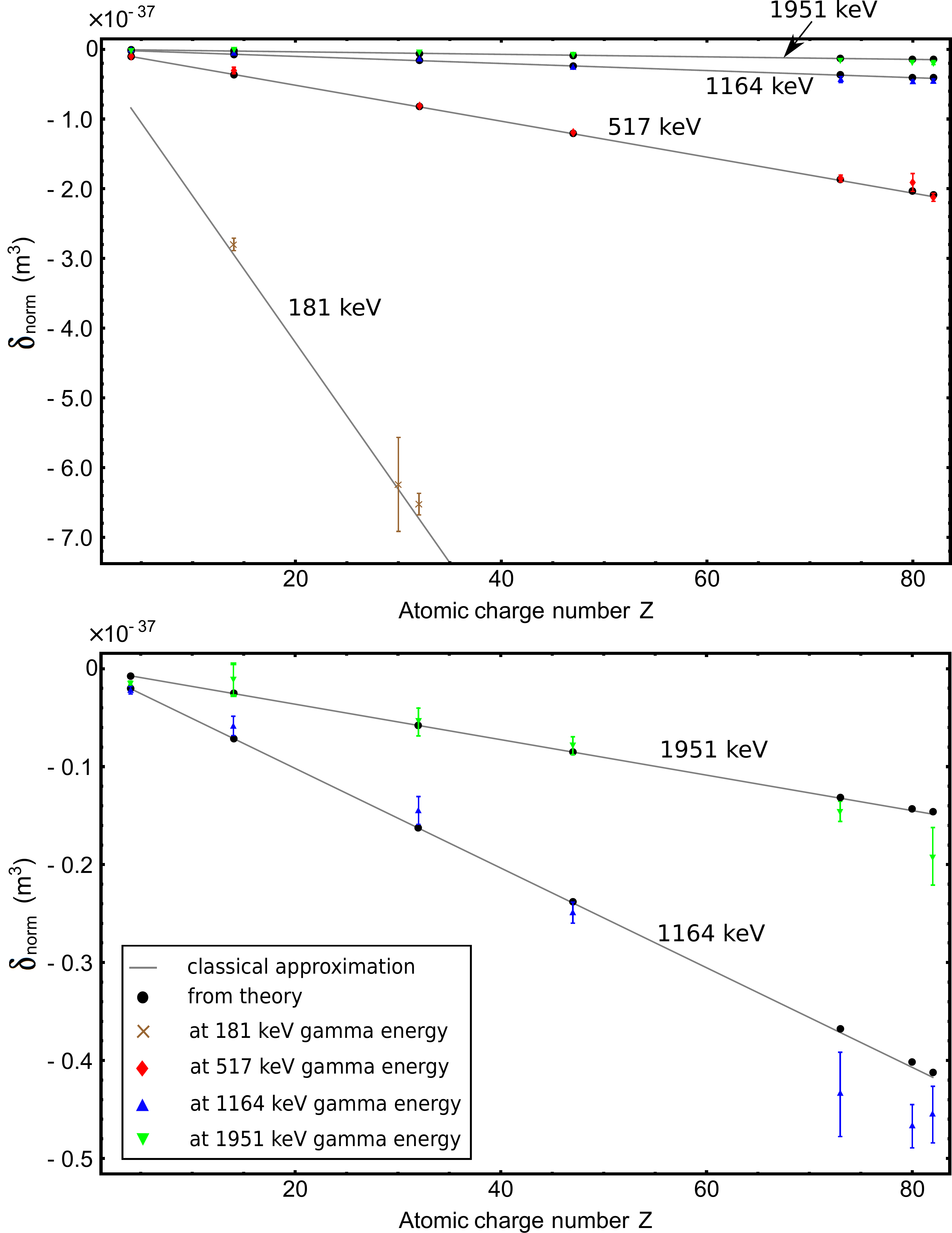}
\caption{\label{fig2}(color online) The normalized real part $\delta_{\rm norm}$ of the complex index of refraction as a function of the atomic charge number $Z$. The $Z$ dependence is shown for different photon energies $E_{\gamma}$ (181 keV, 517 keV, 1164 keV, and 1951 keV). A slight deviation of $\delta_{\rm norm}$ from the classical theory is observed at energies above 517 keV. The black dots are calculated $\delta_{\rm norm}$\textit{'s} from theory. The gray dotted line is an interpolation of the classical approximation over the measured atomic charge range.}
\end{figure*}

We have measured the energy dependence of the refractive index for different materials within the energy range from 181 keV to 1951 keV. To allow for better comparison, we have normalized $\delta$ to the atom number density, i.e., $\delta_{\rm norm} = \delta / N_{\rm atom}$. Fig.~\ref{fig1} shows the measured dispersive behavior of the normalized real part of the refractive index $\delta_{\rm norm}$ as dots.  Furthermore, we have included a dispersion curve model according to the classical approximation, drawn by the black solid lines. We observed that the scattering from low-$Z$ materials is fairly well described by the classical Thomson approximation. However, for high-$Z$ materials and high $\gamma$-ray energies $E_\gamma$, we found a deviation from the classical Thomson description. For a better understanding of the $Z$ dependence of the observed effect and to compare with theoretical calculations based on atomic scattering processes, we plot the normalized refractive index $\delta_{\rm norm}$ as a function of atomic charge number $Z$ of the prism material in Fig.~\ref{fig2}. Here, the experimental and theoretical results are shown for three different energies (517 keV, 1164 keV, and 1951 keV). In case of 181 keV there is good agreement with the classical approximation.

The theoretical results shown in Fig.~\ref{fig2} are the combined atomic elastic scattering contributions of Rayleigh, nuclear Thomson, Delbr\"uck, and nuclear giant dipole resonance scattering, discussed in details in the previous section. In the case of lead the theoretical as well as the measured deviation from the classical approximation is at 517 keV: theo. $-$1.1\%, exp. (0.4$\pm$2.7)\%; at 1164 keV: theo. $-$1.3\%, exp. (9$\pm$7)\%; at 1951 keV: theo. $-$1.4\%, exp. (29$\pm$20)\%. In Table~\ref{table1} we show the individual contributions of atomic interaction processes to the real part of the forward scattering amplitude at different energies for two examples, a low-$Z$ material (beryllium, Be) and a high-$Z$ material (lead, Pb). The classical Thomson amplitude does not depend on the photon energy and linearly increases with $Z$, while the relativistic correction to the Rayleigh amplitude as well as the Delbr\"uck term increase with a growth of the photon energy and nuclear charge $Z$. At high energy and high $Z$ the Delbr\"uck and the relativistic Rayleigh contribution to the scattering amplitude lead to a slight increase of the total scattering amplitude.
\begin{table}[]
  \centering
  \caption{\label{table1} Individual contributions of the atomic elastic scattering processes to the real part of the forward scattering amplitude $Re A(E_\gamma)$ for Be and Pb. The total experimental and theoretical values of $Re A(E_\gamma)$ are given for a comparison in units of the classical electron radius $r_e$.}
\begin{ruledtabular}  
\begin{tabular}{cccc}
Contribution  & 517 keV  & 1164 keV & 1951 keV \\ \hline
Be ( $Z = 4$ )& & &                 \\
&&& \\
Classical        & 100.00\% & 100.00\% & 100.00\% \\
rel. Rayleigh &$-$0.02\% &$-$0.02\% &$-$0.02\% \\
Delbr\"uck    &   0.00\% &   0.00\% &   0.00\% \\ 
nucl. Thomson &   0.02\% &   0.02\% &   0.02\% \\
nucl. reson.  &   0.00\% &   0.00\% &   0.00\% \\ \hline
$Re A(E_\gamma)$ theo. &$-$ 4.0    &$-$ 4.0    &$-$ 4.0      \\ 
$Re A(E_\gamma)$ exp.&$-$ 3.9(2) &$-$ 4.3(7) &$-$ 7.3(1.0) \\ \hline \hline
Pb ( $Z = 82$ )& & &                \\
&&& \\
Classical        & 101.12\% & 101.28\% & 101.45\% \\
rel. Rayleigh &$-$1.13\% &$-$1.22\% &$-$1.22\% \\
Delbr\"uck    &$-$0.01\% &$-$0.08\% &$-$0.25\% \\
nucl. Thomson &   0.02\% &   0.02\% &   0.02\% \\
nucl. reson.  &   0.00\% &   0.00\% &   0.00\% \\ \hline
$Re A(E_\gamma)$ theo. &$-$81.1      &$-$81.0      &$-$ 80.8  \\ 
$Re A(E_\gamma)$ exp.&$-$82.3(2.2) &$-$89.4(5.7) &$-$106(16) \\
\end{tabular}
\end{ruledtabular}
\end{table}

The calculated atomic elastic scattering amplitudes explain the experimental results for low energies and especially for low-$Z$ materials. For high energies and high $Z$ there is a slight deviation between the calculated and experimentally determined real part of the forward scattering amplitude. Furthermore, at 1951 keV there is a deviation for beryllium ($Z = 4$). We attempted to explain the deviation by additional nuclear resonances. In particular, the $^9$Be nucleus has a low-lying $1/2+$ resonance at an energy of 1.735 MeV and with a large neutron-decay width of 225 keV \cite{sumiyoshi:2002:467}. Employing the dispersion relation and experimental photoneutron cross section \cite{sumiyoshi:2002:467}, we calculate the corresponding contribution to the scattering amplitude to be $2 \times 10^{-4}\, r_e$ at an energy of 1951 keV, wand therefore cannot explain the measurement. Thus, the deviation between theory and experiment in the case of beryllium remains unexplained. On the one hand the very small magnitude of the dispersive effects in Be at these high energies make the measurement more susceptible to very small non-linear drifts not compensated by our measurement scheme. To investigate this issue further, we suggest 
performing measurements of the angular-differential cross section for the
elastic scattering process at the same photon energy.

Overall, the measured refractive indices are in good agreement with the
theoretical values. In the worst case of high-$Z$ materials and at photon
energies above 1 MeV, the deviation does not exceed $2\sigma$, where $\sigma$
is the experimental uncertainty of one standard deviation. Furthermore, we have shown theoretically and
confirmed experimentally that the refractive index at MeV $\gamma$-ray energies
is well described by the optical relation and the classical Thomson amplitude
for a wide range of materials. For the cases considered, the corrections to
the classical amplitude appear to be relatively small and reach their maximum
value of about 1.5\% for Pb at 1951 keV photon energy.

\section{Discussion of the experimental results for applications}
\begin{figure}
\includegraphics[width=0.47\textwidth]{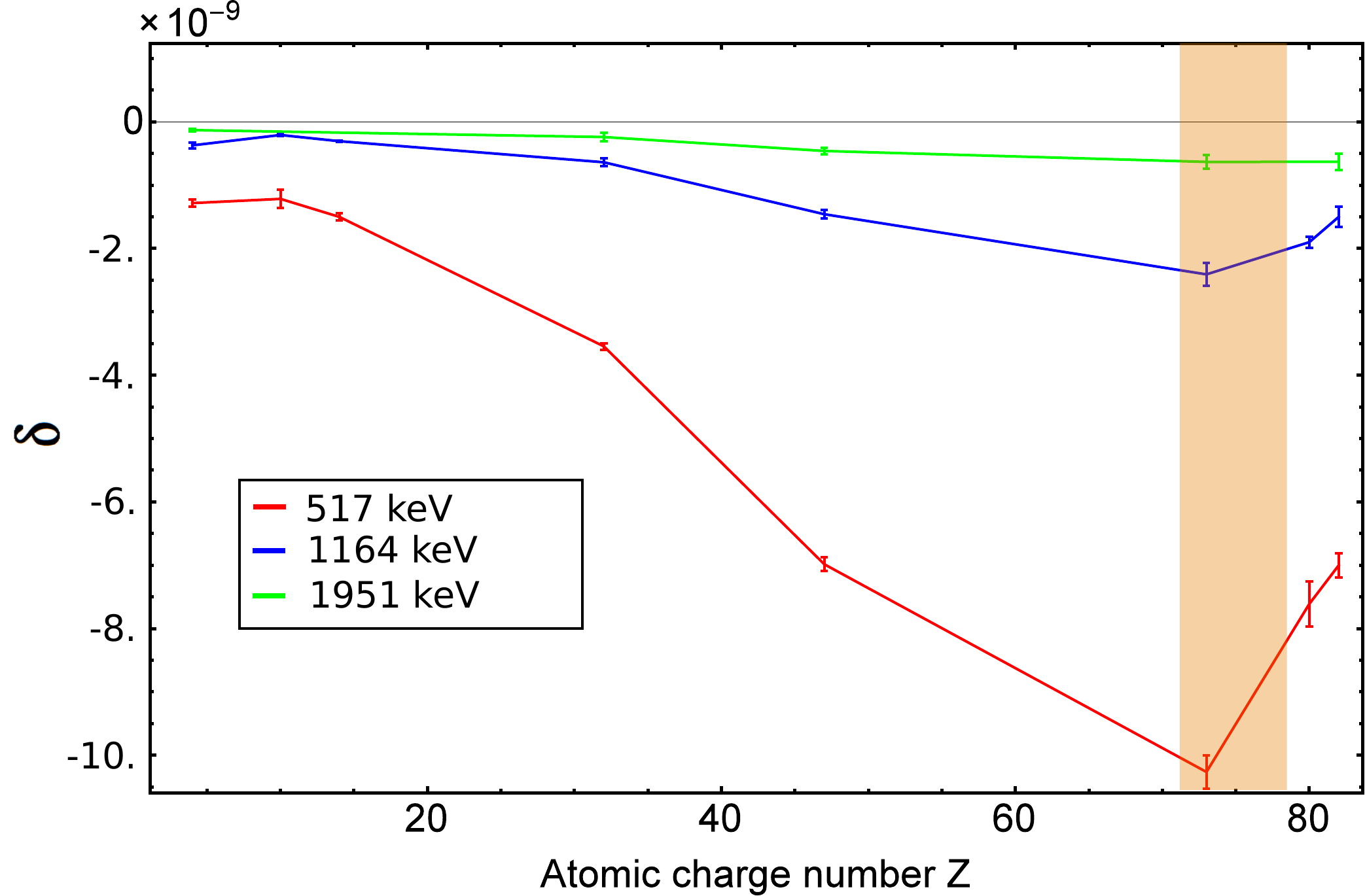}
\caption{\label{fig3}(color online) $Z$ dependence of the measured $\delta$ for three different photon energies. The shaded area highlights the range of materials with maximum mass density.}
\end{figure}
\begin{figure}
\includegraphics[width=0.47\textwidth]{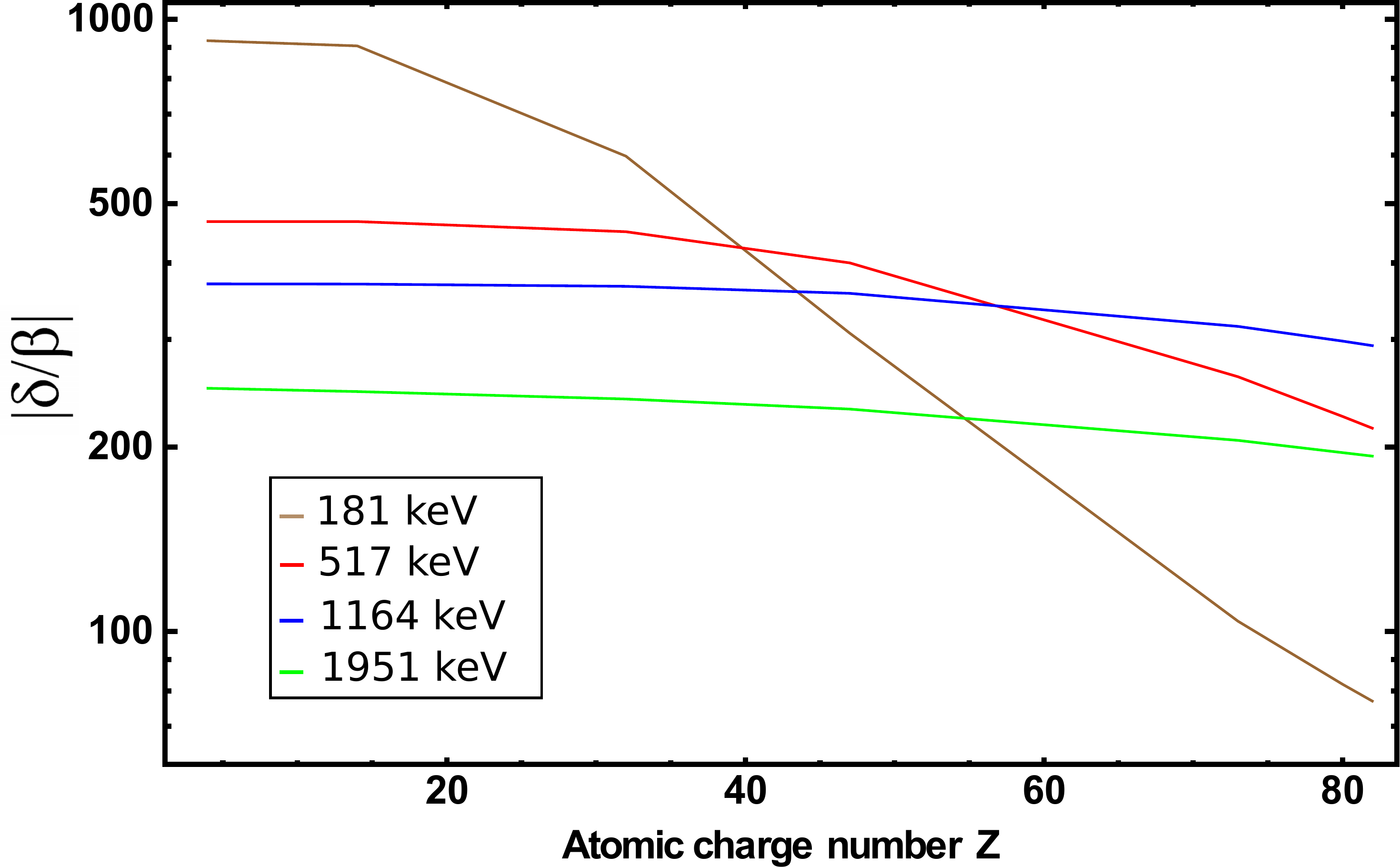}
\caption{\label{fig4}(color online) $Z$ dependence of the relation between $\delta$ (refractive power) and $\beta$ (absorption) for different photon energies. This Figure-of-Merit (FOM) demonstrates that high-$Z$ materials more suitable for $\gamma$-ray refractive optics compared to low-$Z$ materials as there is no significant penalty in of the FOM while the much higher value of $\delta$  allows for  compact focusing lens arrays. }
\end{figure}
Beside the fundamental physics question of the forward scattering behavior during $\gamma$-ray interactions with matter, a further question is the feasibility of refractive $\gamma$-ray optics. In the past the focusing of X-rays was successfully demonstrated, which is today established in X-ray imaging applications. X-ray refractive lenses were realized up to 200 keV photon energies \cite{lengeler1998transmission,0022-3727-38-10A-042}. At such energies the lens systems were calculated via the classical approximation. In the near future, novel highly brilliant and high-energy $\gamma$-ray sources, like ELI-NP \cite{ELINP}, will be realized. The success of a number of different applications in nuclear physics might be boosted by the capability of $\gamma$-beam focusing and collimation. At such high energies conventional focusing optics based on diffraction, like mirrors, multilayer Laue lenses or Fresnel zone plates \cite{ice2011race}, are very difficult to operate, mostly because the angular acceptance range for diffraction decreases strongly with photon energy. This is less relevant for refractive optics. Here, the main problem is the fast convergence of the value of the refractive index towards unity. Based on our investigation of the refractive behavior of several different materials at $\gamma$-ray energies we show in Fig.\ref{fig3} the real part of the measured refractive index $\delta$ as a function of the atomic charge number $Z$. The results demonstrate that there is a range where the refractive index becomes maximal for all energies, which means that these materials are suitable candidates for refractive optics. For those materials we have measured a $\delta$ from -1.2$\times$10$^{-8}$ to -2.5$\times$10$^{-9}$ between 517 keV and 1164 keV. However, for high-density materials we have to consider that absorption attenuates the photon beam. Therfore, it is important to define a Figure-of-Merit (FOM) for a refractive lens system. In that case we write as FOM the relation of the refractive power $\delta$ to the absorption of the lens material given as the imaginary part of the complex refractive index. The imaginary part $\beta$ defines the energy dependent dissipative part and is written as $\beta(E_{\gamma})=\sigma_{\text{tot}}(E_{\gamma})(\hbar c)N_{\text{atom}}/2E_{\gamma}$; where $\sigma_{\text{tot}}$ is the total photo absorption cross section and $N_{\text{atom}}$ the atom number density. In Fig.\ref{fig4} we show the FOM as a function of the atomic charge number $Z$ for different $\gamma$-ray energies. For the calculation of $\beta(E_{\gamma})$ we used the total photon absorption cross section from the XCOM database of the National Institute of Standards and Technology (NIST) \cite{XCOM}. Figure \ref{fig4} shows that high-$Z$ materials are more suitable for refractive optics at high $\gamma$-ray energies compared to low-$Z$ materials. With such materials it is feasible to realize refractive focusing optics within the high photon energy range above 500 keV while keeping the overall number of lens elements manageable due to the higher value of $\delta$. For the lower energies in our study and X-ray energies in general, high-density materials are not suitable because absorption is very high and scattering processes like Compton scattering lead to degradation of the focusing power. However, at high photon energies, especially around 1 MeV, the detrimental effect of  Compton scattering decreases compared to Rayleigh scattering and other coherent scattering processes. As shown in the past, refractive X-ray lenses can be realized by a series of lenses, so-called compound refractive lens (CRL) systems. In a further work, we will describe in detail newly developed lens systems for the $\gamma$-ray energy range. These lens systems are made of high density, high $Z$ materials with a design with less absorption than X-ray refractive lenses.  Besides the focal length, the effective aperture as well as the transmission are important values in lens development. The focal length is calculated by the lens maker formula $f=R/(2N\delta)$, where $R$ is the radius of curvature of the lens surfaces, $N$ is the number of lenses and $\delta$ is the refractive power. The effective aperture describes the geometric aperture reduced by absorption and Compton processes, and is given by $D_{\text{eff}}=2R_0\sqrt{\frac{[1-\exp(-\mu Nl)]}{\mu Nl}}$; $\mu$ is the photon attenuation coefficient, $l$ the half length of each lens within the lens system and $2R_0$ is the geometric aperture. The transmission, depending on the absorption and Compton scattering, refers to the fraction of transmitted intensity in comparison to the intensity, which is reduced by the geometric aperture, only. The expression is $T=\frac{1}{2\mu N l}[1-\exp(-2\mu N l)]$. In summary, for applications like the development of refractive optics for $\gamma$ rays with an energy up to 1 MeV, the classical approximation for the calculation of the refractive index can be used. For higher $\gamma$-ray energies, above 1 MeV, the optical behavior has to be investigated in more detail as mentioned in chapter IV.

\section{Conclusions}
In an experimental campaign we have investigated the index of refraction using the new ultra high-resolution gamma spectrometer GAMS6 at the ILL in Grenoble (France). We measured the refractive index of several pure, compound and fluid materials (Be, Si, Ge, Ag, Ta, Pb, Hg, SiO$_2$) within a wide range of atomic charge numbers $Z$ ($Z=4$ to $Z=82$). The aim was to investigate the energy as well as the $Z$ dependence of the refractive index and therefore test the classical approximation for the description of the refractive effect.

For high-$Z$ materials and high photon energies, we observed a slight deviation of the real part $\delta$ from the classical approximation, both depending on the photon energy and $Z$. Also, in case of beryllium ($Z=4$) we observed a deviation from theory. We investigated the contribution of nuclear resonances to the forward scattering amplitude with the result that the contribution is too small to account for the observed deviation from theory. Overall, the measured refractive indices are in reasonably good agreement with the theoretical values. We have shown theoretically and confirmed experimentally that the refractive index at MeV $\gamma$-ray energies are well described by the optical relation and the classical Thomson amplitude for a wide range of materials. For the explanation of the deviation between theory and experiment in case of beryllium as well as for the high-$Z$ materials at high $\gamma$-ray energies, it needs further investigations. We suggest to perform measurements of the angular-differential cross section for the elastic scattering processes at the same photon energies. Here the contributions of different scattering physics can be investigated in more detail at the MeV $\gamma$-ray energy range and allow for test of corrections in atomic scattering theory.

Furthermore, from an applied point of view, the classical approximation can be used for $\gamma$-ray energies up to  MeV energies for refractive optics development. We confirmed that high-$Z$ materials with high mass density are the best choice for refractive $\gamma$-ray optics. Based on our results, in an upcoming work we describe in detail the implementation of lens systems under consideration of simulation studies of optical as well as atomic interaction physics.

\nocite{*}

\end{document}